\documentclass[aps,prl,preprint,superscriptaddress,amsmath,amssymb]{revtex4}
\usepackage{graphicx}
\usepackage{color}
\begin{document}
\title{Bond-selective fragmentation of water molecules with intense, ultrafast, carrier envelope phase stabilized laser pulses}
\author{D. Mathur}
\email{atmol1@tifr.res.in}
\affiliation{Tata Institute of Fundamental Research, 1 Homi Bhabha Road, Mumbai 400 005, India}
\affiliation{Department of Atomic and Molecular Physics, Manipal Institute of Technology, Manipal University, Manipal 576 104, India}
\author{K. Dota}
\affiliation{Tata Institute of Fundamental Research, 1 Homi Bhabha Road, Mumbai 400 005, India}
\affiliation{Department of Atomic and Molecular Physics, Manipal Institute of Technology, Manipal University, Manipal 576 104, India}
\author{J. A. Dharmadhikari}
\affiliation{Department of Atomic and Molecular Physics, Manipal Institute of Technology, Manipal University, Manipal 576 104, India} 
\author{A. K. Dharmadhikari}
\affiliation{Tata Institute of Fundamental Research, 1 Homi Bhabha Road, Mumbai 400 005, India}

\begin{abstract}
Carrier envelope phase (CEP) stabilized pulses of intense 800 nm light of 5 fs duration are used to probe the dissociation dynamics of dications of isotopically-substituted water, HOD. HOD$^{2+}$ dissociates into either H$^+$ + OD$^+$ or D$^+$ + OH$^+$. The branching ratio for these two channels is CEP-dependent; the OD$^+$/OH$^+$ ratio (relative to that measured with CEP-unstabilized pulses) varies from 150\% to over 300\% at different CEP values, opening prospects of isotope-dependent control over molecular bond breakage.  The kinetic energy released as HOD$^{2+}$ Coulomb explodes is also CEP-dependent. Formidable theoretical challenges are identified for proper insights into the overall dynamics which involve non-adiabatic field ionization from HOD to HOD$^+$ and, thence, to HOD$^{2+}$ via electron rescattering. 
\end{abstract}
\maketitle

Femtochemistry experiments have, in the course of a few decades, led to qualitative leaps in our understanding of chemical reactivity and molecular dynamics at a microscopic level \cite{zewail}. The dielectics governing femtochemistry are based on Born-Oppenheimer potential energy surfaces (PES): the molecule's electronic wave function adiabatically adapts to the nuclear dynamics that occur on the PES while, concomitantly, the nuclear dynamics are a consequence of the forces generated on the PES that describes how the molecule's electronic energy varies with nuclear geometry. The time evolution of a chemical reaction becomes amenable to control by application of a femtosecond laser pulse whose frequency and amplitude are decided upon by an experimentalist employing pulse shaping methods \cite{pulseshaping}. The recent  availability of intense laser pulses whose durations are short enough to accommodate only a couple of optical cycles has, however, introduced a new paradigm: molecular dynamics might be amenable to control by an experimentalist who manipulates not the instantaneous frequency of the laser pulse but its electric field waveform. The parameter of importance then becomes the carrier envelope phase (CEP), which quantifies the temporal offset between the maximum of a laser pulse's envelope and the maximum of the optical cycle. Using CEP-stabilized pulses, molecules can be irradiated such that, at fixed intensity, the magnitude of the field experienced by the molecule is experimentally varied via control of the CEP. Intense few-cycle pulses within which the optical field is precisely fixed via CEP control provide a fillip to attosecond science in that they offer unprecedented opportunities for experimentalists to be able to control the moment of ``birth" of an electron wave packet (in the tunnel ionization process) as well as its subsequent motion (in the rescattering of the electron wave packet due to the oscillating ponderomotive potential). Initial experiments with CEP-stabilized pulses have already begun to reveal new facets of strong field ionization, such as suppression of non-sequential ionization, time-dependent bond hardening \cite{Dota} and the role of inner valence orbitals in rescattering-driven fragmentation of organic molecules \cite{Xie}. Might it be possible to selectively break one of two bonds in a simple triatomic molecule like water? We focus here on experiments that seek an answer to this question. We explore field-induced ionization and fragmentation of the water molecule (in which we have replaced one of the H-atoms by a D-atom). Our optical field is strong enough to form HOD$^{2+}$; we explore the possibility of this dication preferentially dissociating into either H$^+$ + OD$^+$ (breaking the O-H bond) or D$^+$ + OH$^+$ (breaking the O-D bond). We show that, at fixed laser intensity, CEP affects the propensity with which one or the other bonds breaks. Our results present a formidable challenge to theory as a proper quantal description of the dissociation dynamics demands {\em time-dependent} computations of the field-distorted potential energy surfaces of the dication. In the absence of such surfaces we rationalize our experimental observations by considering wavepacket propagation on surfaces of HOD$^{+}$ and subsequent rescattering-driven transition to surfaces of HOD$^{2+}$: we qualitatively show that the observed selectivity of bond breaking is simply related to isotopic mass dependent wavepacket propagation on such surfaces. Our model predicts a CEP dependence for the kinetic energy released (KER) upon dissociation of HOD$^{2+}$; our measurements confirm that the energetics accompanying bond breakages is, indeed, CEP-dependent. Such dependence indicates that KER spectra might be a sensitive monitor of how ultrafast molecular dynamics is affected by the instantaneous magnitude of the time-varying field and how effectively such dynamics might be amenable to control. Formidable theoretical challenges are identified that need to be overcome if proper insights are to be developed into the overall dynamics.

Recent molecular dynamics work with CEP-unstabilized few-cycle pulses \cite{CS2paper} has provided evidence that molecular ionization dominates the ionization spectrum, with hardly any fragmentation. This is most likely a consequence of multiple ionization being mostly due to non-sequential (NS) ionization: simultaneous tunneling of more than one electron through field-distorted potential functions.  In contrast, experiments with CEP-controlled 2-cycle pulses have shown that atomic fragmentation of molecules is enhanced, and that it depends on the instantaneous field strength \cite{CEP_CS2}. Moreover, such enhancement is at the expense of molecular ionization \cite{CEP_CS2}. Yamakawa {\it et al.} \cite{Yamakawa} have shown suppression of NS ionization in the few-cycle regime and recent results on multiple ionization of Xe have established the phase dependence of NS ionization and its contribution to the formation of highly charged ions; the relative yields of such ions show CEP-dependent behavior that highlights the field-dependence (not intensity-dependence) of NS ionization. The major driver of NS ionization is rescattering \cite{Bhardwaj} and it is, therefore, not unexpected that NS-ionization should be CEP-dependent because it is CE-phase that determines when the ionized electron is ``born" in the course of a laser pulse. 
      
There is a paucity of work on molecular ionization with CEP-stabilized pulses; in earlier work \cite{Xie,Leone} pulse intensities were limited to $\sim$10$^{13}$ W cm$^{-2}$ where ionization is due to both tunnel ionization and multiphoton ionization. The former is a field-dependent processes while the latter depends on the intensity envelope of the laser pulse. In contrast, we have chosen to make measurements in the 10$^{15}$ W cm$^{-2}$ range (typical contrast ratio $\sim$10$^6$). Ionization of HOD is, therefore, by tunneling. Our experiments were conducted using a Ti:Sapphire oscillator (78 MHz repetition rate) whose pulses were first amplified (in a 4-pass amplifier) and then stretched (to $\sim$200 ps). The pulse shape and duration were controlled with an acousto-optic dispersive filter. The resulting pulses, down-converted to 1 kHz repetition rate by an electro-optical modulator, were amplified (in a 5-pass amplifier) and compressed. These pulses were typically 22 fs long; further compression, to 5 fs, was accomplished using a 1 m-long Ne-filled hollow fiber and chirped dielectric mirrors. CEP stabilization was achieved using a fast-loop in the oscillator and a slow-loop in the amplifier \cite{krause}. As shown in our recent work \cite{Dota} on time-dependent bond-hardening in Si(CH$_3$)$_4$ at CEP=0, we typically obtain phase jitter $<$110 mrad over the course of measurements, as determined by an $f-2f$ interferometer at 1 kHz spectrometer acquisition rate with 920 $\mu$s integration time and 84 ms loop cycle. With CEP-stabilization, laser energy stability of 0.4\% rms was readily achieved. Linearly-polarized pulses were transmitted to our molecular beam apparatus through a thin (300 $\mu$m) fused-silica window \cite{earlier}. Our laser beam was focused to 7 $\mu$m (width at 1/e$^2$) using a 5 cm curved mirror placed within our ultrahigh vacuum chamber. We used a linear time-of-flight spectrometer to monitor ionization (with unit collection efficiency) by acquiring data (at 1 kHz) in list mode using a segmented-mode 2.5 GHz oscilloscope. We established intensity values by measuring Xe-ionization spectra and calibrating with the appearance threshold of Xe$^{6+}$ \cite{CEP_CS2}. HOD was formed by admixing equal volumes of H$_2$O and D$_2$O and monitoring the formation of HOD via absorption spectroscopy \cite{method}.   

\begin{figure}
\includegraphics[width=8cm]{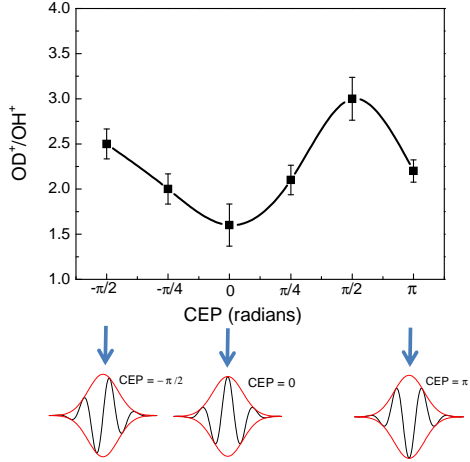}
\caption{(Color online) CEP-dependence of the OD$^+$/OH$^+$ ratio relative to the value measured using CEP-unstabilized pulses. The lower part of the figure shows optical field variation within a single pulse at specified CEP values.}
\end{figure}

Our measurements focused on the dissociation of the HOD$^{2+}$ dication into either of the two channels: H$^+$ + OD$^+$ or D$^+$ + OH$^+$. This is the ionic analog of the reaction HOD$\rightarrow$H + OD or D + OH, which has been used by theorists as a prototype for direct dissociation on a single, isolated potential energy curve \cite{tiwari}. In our case, the branching ratio for the two channels is 0.5 when the dication state is accessed using CEP-unstabilized pulses of 1$\times$10$^{15}$ W cm$^{-2}$ intensity: the measured OD$^+$/OH$^+$ ratio is consistent with the relative yields of HOD, D$_2$O, and H$_2$O in our mixture of light and heavy water, as determined by absorption spectroscopy \cite{method}. Typical raw data showing H$^+$ + OD$^+$ or the D$^+$ + OH$^+$ ion pairs from HOD$^{2+}$ precursors, as well as analogous ion pairs from H$_2$O$^{2+}$ and D$_2$O$^{2+}$ are shown in the Supplemental Material \cite{method}. The OD$^+$/OH$^+$ ratio alters dramatically when we use CEP-stabilized pulses (of fixed peak intensity) to initiate ionization. Figure 1 shows the strong CEP-dependence of the OD$^+$/OH$^+$ ratio {\em relative to} that measured with CEP-unstabilized pulses. The ratio varies over the range $\sim$1.5 at CEP=0 to as much as $\sim$3 for CEP=+$\pi$/2. We reiterate that both OD$^+$ and OH$^+$ are constituents of ion-pairs formed upon Coulomb explosion of the HOD$^{2+}$ dication. Some typical field variations within a single pulse are also schematically shown in Fig. 1 for different values of CEP. As our pulse comprises barely two optical cycles, strong field ionization of HOD occurs in the first half cycle. The ionized electron is driven back towards HOD$^+$ as the sign of the optical field reverses. The resulting recollision then ionizes HOD$^+$ to a dication state, HOD$^{2+}$, which Coulomb-explodes into H$^+$ + OD$^+$ or D$^+$ + OH$^+$. This two-step sequence of events essentially accounts for the CEP-dependent bond-breakage that we observe. The initial field ionization event gives rise to either H$^+$ + OD or D$^+$ + OH. In the former case the wavepacket propagates further along the HOD$^+$ PES than in the latter. Consequently, upon rescattering, the upward transition occurs to different locations of the dication PES (Fig. 2). As the ``birth" of the electron that initiates rescattering is determined by the value of the optical field experienced by the HOD molecule within a single pulse, it is natural that CEP values determine the overall Coulomb explosion dynamics.

\begin{figure}
\includegraphics[width=6cm]{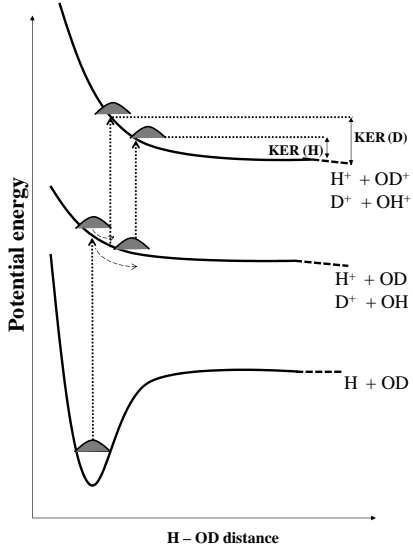}
\caption{(Color online) Schematic potential energy (PE) curves of HOD, HOD$^+$, and HOD$^{2+}$. The strong optical field induces tunnel ionization into HOD$^+$ (single vertical arrow). As the nuclear wavepacket propagates along the HOD$^+$ PE curve, electron rescattering induces a transition to the HOD$^{2+}$ PE curve (two vertical arrows depicting the isotopically dependent upward transition. Dissociation along the dication PE curve yields isotopically dependent values of kinetic energy release (KER). Note that the PE curves do not account for either the optical field or its time dependence.}
\end{figure}

Our model is simplistic, as discussed in the following, but it does provides us with an opportunity of making an experimental test of its robustness in a qualitative sense and, concomitantly, exposing the shortcomings of the simple picture that underpins the model. As seen in Fig. 2, the two different wave packets on the dication PES would be expected to give rise to different amounts of kinetic energy released (KER) in the center-of mass frame as the HOD$^{2+}$ dication dissociates. If the shapes of the PE curves remained static in the course of a single pulse, it would be expected that KER(D), denoting the kinetic energy release accompanying formation of D$^+$ + OH$^+$, would have a higher value than KER(H), the kinetic energy release accompanying formation of H$^+$ + OD$^+$.  However, our measurements (Fig. 3) reveal that the maximum value of KER(D) ($\sim$10 eV) is significantly lower than that for KER(H) ($\sim$20 eV). For CEP-unstabilized pulses, KER(H) had a maximum value of 23.4 eV while the corresponding value for KER(D) was only 4.2 eV. Similarly for CEP=0, KER(H) has a higher value than KER(D) (17.2 eV and 9.2 eV, respectively). These measurements manifest the reality that the PE curves do {\em not} remain static in the course of our ultrashort laser pulse: with two-cycle pulses the electronic and nuclear motions are very strongly coupled and, consequently, a coupled electron-nuclear Schr\"odinger equation needs to be solved in order to obtain proper, time-dependent, field-dressed PE curves. This remains a formidable technical challenge, one that has been taken up only for H$_2^+$ \cite{TDPES} for 2-8 fs pulses.  The results are counter-intuitive in that they show the potential well in the lowest-energy H$_2^+$ state collapsing as the laser pulse reaches its peak amplitude of {\em ca.} 10$^{15}$ W cm$^{-2}$, only to regain its form in the trailing edge of the pulse. The PE curves show substantial fluctuations in shape within the ultrashort duration of a single pulse. Such fluctuations would also occur in HOD and its ions. The measured KER values would, of course, not only reflect the fluctuating shapes of HOD$^+$ PE curves but also the value of CEP that determines the rescattering dynamics that, in turn, govern the HOD$^+\rightarrow$HOD$^{2+}$ transition. Consequently, we would expect measured KER values to also depict a strong CEP-dependence. Figure 3 shows that this is, indeed, the case. 

\begin{figure}
\includegraphics[width=10cm]{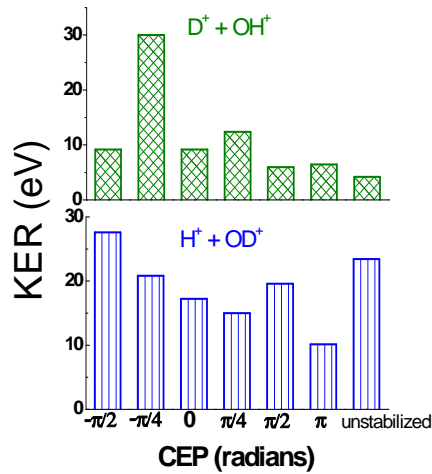}
\caption{(Color online) CEP-dependent values of the maximum kinetic energy released (KER) as HOD$^{2+}$ dissociates into two ion pairs. KER values obtained with CEP-unstabilized pulses are also shown.}
\end{figure}

Attempts to correlate our experimental observations with wavepacket dynamics on time-dependent, field-dressed, coupled PE curves also reveal another facet of strong field molecular science: how valid is it to consider the few-cycle dynamics in terms of molecular orbitals that are implicitly based on an adiabatic, single-active-electron picture? The adiabaticity in the electron dynamics in conventional femtochemistry experiments originates in the fact that, upon reaching the field strength required for tunneling, the electron has sufficient time to ionize before there is any further increase in field strength. In our experiments, the electronic response ceases to be adiabatic as HOD can ``survive" to higher laser intensities before ionization occurs. Hence, the ionized electron is exposed to much higher, rapidly increasing field strengths, as a result of which it gains more energy before the onset of rescattering. There are analogies in high harmonic generation (HHG) from atomic gases like Ar \cite{Christov} where the HHG spectrum measured with 25 fs pulses contains many more harmonics than with 100 fs pulses of equal intensity. This is a consequence of Ar in an ultrashort field surviving to higher laser intensities, a consequence of the non-adiabatic response of the atomic dipole to the rise time of the ultrashort pulse \cite{Christov}. We visualize the breakdown of adiabaticity in our few-cycle field ionization in the following terms. If $E_a$ is the field amplitude that distorts the HOD PES to the level of the ionization energy, IE, early work on strong field ionization \cite{Augst} established that $E_a=IE^2/4$. Hence, the ratio of the tunneling time to the laser period (Keldysh parameter, $\gamma$) becomes $\gamma = 2E_a^{0.25}\omega/E$. For ionizing HOD (IE=12.6 eV) at a field strength of $E$=$E_a/2$, the ionized electron would take about the same time to tunnel through the field-distorted potential barrier as the optical period (2.7 fs) of 800 nm light. With $\gamma\sim$0.3, we expect the electron wavepacket to experience a strongly delayed response vis-a-vis our field. Is it possible to reconcile this phase lag with the tunneling picture that is central to our ionization dynamics? The tunneling ionization rate within the oft-used ADK (Ammosov-Delone-Keldysh) formula \cite{ADK} is a function of the instantaneous value of the field envelope such that, after integration over time, is directly proportional to the pulse duration. In the limit of weak light fields, this is, of course, consistent with the Fermi Golden Rule: the ionization probability scales with photon number (the pulse energy being the experimental manifestation) until, at high enough laser intensities, the ionization saturates. For our ultrashort pulses, this adiabatic picture naturally breaks down as ionization occurs within a single optical cycle (Fig. 1). Thus, our results present a not-inconsiderable challenge to develop an adequate nonadiabatic framework within which strong-field molecular ionization can be understood in the few-cycle regime.  

The challenge does not end there. Field-induced ejection of an electron from the nonbonding outermost 1$b_1$ orbital yields the $^2B_1$ ground electronic state of HOD$^+$. Similarly, electron ejection from the 3$a_1$ orbital yields the cation's first excited state, $^2A_1$. However, the notion of handling the ionization dynamics just one electron at a time seems invalid under our experimental conditions. The trailing edge of our laser pulse can strongly couple closely-lying PE surfaces, leading to a superposition of states and consequent localization of electron wavepackets. In the case of H$_2^+$, experimental evidence has already been obtained for such electron wavepacket localization, with the site and extent of localization being amenable to CEP-control \cite{localization}. Hence, concurrently with the need to develop a nonadiabatic theory of strong field ionization is the requirement to be able to adequately deal with multielectron excitation events which forms molecular ions in highly excited states.  Apart from our experiments, other evidence has been obtained of the breakdown of adiabatic, single-active-electron dynamics in molecules in studies of the wavelength dependence of ionization and fragmentation of hydrocarbon molecules \cite{lezius}; multielectron excitation results in formation of molecular ions in highly excited states.  Unexpectedly high KERs accompany dissociation of such states \cite{ultrafast_methane}, as is seen in Fig. 3. 

Financial support from the Department of Science and Technology is acknowledged by JAD (Women Scientists Scheme) and DM  (J. C. Bose National Fellowship).


\begin{thebibliography}{00}
\bibitem{zewail}A. H. Zewail, Angew. Chem. Int. Ed. {\bf 39}, 2587 (2000).
\bibitem{pulseshaping}G. Vogt, G. Krampert, P. Niklaus, P. Nuernberger, and G. Gerber, Phys. Rev. Lett. {\bf 94}, 068305 (2005), C. Daniel, {\it et al.}, Science {\bf 229}, 536 (2003), H. Rabitz, R. de Vivie-Riedle, M. Motzkus, and K. Kompa, Science {\bf 288}, 824 (2000), and references therein.
\bibitem{Dota}K. Dota, M. Garg, A. K. Tiwari, J. A. Dharmadhikari, A. K. Dharmadhikari, and D. Mathur, Phys. Rev. Lett. {\bf 108}, 073602 (2012).
\bibitem{Xie}X. Xie, {\it et al.}, Phys. Rev. Lett. {\bf 109}, 243001 (2012).
\bibitem{CS2paper}D. Mathur, A. K. Dharmadhikari, F. A. Rajgara, and J. A. Dharmadhikari, Phys. Rev. A. {\bf 78}, 013405 (2008).
\bibitem{CEP_CS2}D. Mathur, K. Dota, A. K. Dharmadhikari, and J. A. Dharmadhikari, Phys. Rev. Lett. {\bf 110}, 083602 (2013). 
\bibitem{Yamakawa}K. Yamakawa, {\it et al.}, Phys. Rev. Lett. {\bf 92}, 123001 (2004).
\bibitem{Bhardwaj}V. R. Bhardwaj, {\it et al.}, Phys. Rev. Lett. {\bf 86}, 3522 (2001).
\bibitem{Leone}M. J. Abel, {\it et al}, J. Phys. B {\bf 42}, 075601 (2009).
\bibitem{krause}J. Rauschenberger {\it et al.}, Laser Phys. Lett. {\bf 3}, 37 (2006).
\bibitem{earlier}D. Mathur, A. K. Dharmadhikari, F. A. Rajgara, and J. A. Dharmadhikari, Phys. Rev. A {\bf 78}, 023414 (2008), A. K. Dharmadhikari, J. A. Dharmadhikari, F. A. Rajgara, and D. Mathur, Opt. Express {\bf 16}, 7083 (2008), F. A. Rajgara, D. Mathur, A. K. Dharmadhikari, and C. P. Safvan, J. Chem. Phys. {\bf 130}, 231104 (2009).
\bibitem{method}H$_2$O+D$_2$O$\rightarrow$2HOD. See Supplemental Material for a typical absorption spectra of a mixture of H$_2$O and D$_2$O, showing stretching and bending vibrational modes that indicate formation of HOD. Also shown are coincidence maps depicting formation of various ion pairs from HOD$^{2+}$, H$_2$O$^{2+}$, and D$_2$O$^{2+}$ precursors. 
\bibitem{tiwari}A. K. Tiwari, K. B. M\/oller, and N. E. Henriksen, Phys. Rev. A {\bf 78}, 065402 (2008).
\bibitem{TDPES}M. Garg, A. K. Tiwari, and D. Mathur, J. Phys. Chem. A {\bf 116}, 8762 (2012).
\bibitem{Christov}I. P. Christov, J. Zhou, J. Peatross, A. Rundquist, M. M. Murnane, and H. C. Kapteyn, Phys. Rev. Lett. {\bf 77}, 1743 (1996).
\bibitem{Augst}S. Augst, D. Strickland, D. D. Meyerhofer, S. L. Chin, and J. H. Eberly, Phys. Rev. Lett. {\bf 63}, 2212 (1989).
\bibitem{ADK}M. V. Ammosov, N. B. Delone, and V. P. Krainov, Sov. Phys. {\bf 64}, 1191 (1986). 
\bibitem{localization}M. F. Kling, {\it et al}, Science {\bf 312}, 246 (2006), M. Kremer, {\it et al}, Phys. Rev. Lett. {\bf 103}, 213003 (2009), I. Znakovskaya, {\it et al}, Phys. Rev. Lett. {\bf 108}, 063002 (2012).
\bibitem{lezius}M. Lezius, V. Blanchet, D. M. Rayner, D. M. Villeneuve, A. Stowlow, and M. Yu. Ivanov, Phys. Rev. Lett. {\bf 86}, 51 (2001).
\bibitem{ultrafast_methane}D. Mathur and F.A. Rajgara, J. Chem. Phys. {\bf 124},194308 (2006).
\end{thebibliography}
\end{document}